\documentclass[aps, prl, amsmath, showpacs, preprintnumbers, twocolumn, amssymb, amsmath, superscriptaddress]{revtex4}

\usepackage{graphicx}
\usepackage{mathptmx, textcomp}
\usepackage[latin1]{inputenc}
\usepackage{tabularx}
\usepackage{units}
\usepackage{color}

\begin{document}
\title{Bose-Einstein Condensation of Erbium}
\affiliation{Institut f\"ur Experimentalphysik and Zentrum f\"ur Quantenphysik, Universit\"at Innsbruck, Technikerstra{\ss}e 25, 6020 Innsbruck, Austria}
\author{K. Aikawa}
\affiliation{Institut f\"ur Experimentalphysik and Zentrum f\"ur Quantenphysik, Universit\"at Innsbruck, Technikerstra{\ss}e 25, 6020 Innsbruck, Austria}
\author{A. Frisch}
\affiliation{Institut f\"ur Experimentalphysik and Zentrum f\"ur Quantenphysik, Universit\"at Innsbruck, Technikerstra{\ss}e 25, 6020 Innsbruck, Austria}
\author{M. Mark}
\affiliation{Institut f\"ur Experimentalphysik and Zentrum f\"ur Quantenphysik, Universit\"at Innsbruck, Technikerstra{\ss}e 25, 6020 Innsbruck, Austria}
\author{S. Baier}
\affiliation{Institut f\"ur Experimentalphysik and Zentrum f\"ur Quantenphysik, Universit\"at Innsbruck, Technikerstra{\ss}e 25, 6020 Innsbruck, Austria}
\author{A. Rietzler}
\affiliation{Institut f\"ur Experimentalphysik and Zentrum f\"ur Quantenphysik, Universit\"at Innsbruck, Technikerstra{\ss}e 25, 6020 Innsbruck, Austria}
\author{R. Grimm}
\affiliation{Institut f\"ur Experimentalphysik and Zentrum f\"ur Quantenphysik, Universit\"at Innsbruck, Technikerstra{\ss}e 25, 6020 Innsbruck, Austria}
\affiliation{Institut f\"ur Quantenoptik und Quanteninformation,
 \"Osterreichische Akademie der Wissenschaften, 6020 Innsbruck, Austria}
\author{F. Ferlaino}
\affiliation{Institut f\"ur Experimentalphysik and Zentrum f\"ur Quantenphysik, Universit\"at Innsbruck, Technikerstra{\ss}e 25, 6020 Innsbruck, Austria}

\date{\today}


\begin{abstract}
We report on the achievement of Bose-Einstein condensation of erbium atoms and on the observation of magnetic Feshbach resonances at low magnetic field. By means of evaporative cooling in an optical dipole trap, we produce pure condensates of $^{168}$Er, containing up to $7 \times 10^{4}$ atoms. Feshbach spectroscopy reveals an extraordinary rich loss spectrum with six loss resonances already in a narrow magnetic-field range up to $3$~G. Finally, we demonstrate the application of a low-field Feshbach resonance to produce a  tunable dipolar Bose-Einstein condensate and we observe its characteristic $d$-wave collapse.
\end{abstract}

\maketitle
Ultracold quantum gases have proven to be ideal systems for observing spectacular many- and few-body quantum effects. The large majority of these phenomena rely on the high degree of control over the interparticle interaction achieved with ultracold atoms. In the widely used alkalis, ultracold atoms interact isotropically via a short-range contact potential.
A novel exciting frontier in quantum gas experiments is to access unexplored physical scenarios based on the anisotropic and long-range dipole-dipole interaction (DDI) \cite{Baranov2008tpi,Lahaye2009tpo}.
A dipolar quantum gas is expected to exhibit fascinating phenomena, including novel many-body quantum phases \cite{Goral2000bec, Damski2003coa, Pupillo2008cmp, Jin2011pmi}. The DDI acts in systems having sizable electric or magnetic dipole moments \cite{Lahaye2009tpo}.

In the context of ultracold atomic quantum gases,  pioneering experimental work on strong DDI  has been carried out  with  chromium atoms \cite{Lahaye2007sde,Lahaye2008dwc,Pasquiou2011sra}. Magnetic  lanthanides offer new possibilities for dipolar physics. In such systems, the combination of a large magnetic moment and a large atomic mass leads to a particularly strong dipolar character.  The  demonstration of the first magneto-optical trap of erbium atoms \cite{Mcclelland2006lcw} stimulated growing interest on such species for quantum gas experiments. Very recently, a Bose-Einstein condensate (BEC) and a degenerate Fermi gas of dysprosium have been produced \cite{Lu2011sdb,Lu2012qdd}.
We choose erbium as a promising candidate for experiments on dipolar quantum gases. This species has a number of very appealing features, including a large magnetic moment  $\mu$  of seven times the Bohr magneton, several stable isotopes, a rich energy level scheme \cite{Ban2005lct} with a non-$S$ electronic ground state \cite{note1}, and interesting cold collisional phenomena \cite{Krems2004eia,Connolly2010lsr}.

In strongly magnetic atoms, the competition between the DDI and the contact interaction is very important and gives rise to many intriguing phenomena. The contact interaction is determined by the $s$-wave scattering length $a$ and can be often tuned with external magnetic fields via Feshbach resonances \cite{Chin2010fri}. Tuning of $a$ also controls the balance of these two interactions.
In the case of a novel species in quantum gas experiments, Feshbach resonances and scattering lengths are {\em a priori} unknown. Magnetic lanthanide such as erbium with their large magnetic moments and their non-$S$ electronic ground states present a completely unexplored terrain in ultracold scattering physics. Here the anisotropic interaction is expected to give rise to  novel scattering scenarios, which are not accessible with alkali atoms \cite{Kotochigova2011ait,Petrov2012aif}.

In this Letter, we report on the attainment of  Bose-Einstein condensation of erbium atoms and on the observation of Feshbach resonances in the region of low magnetic fields.  We obtain pure optically trapped BECs of $^{168}$Er containing $7\times 10^4$ atoms. The remarkably high efficiency of evaporative cooling in a standard optical dipole trap indicates favorable scattering properties of the  $^{168}$Er isotope. In addition, the magnetic Feshbach spectroscopy provides first valuable information on the scattering behavior of submerged-shell atoms  at ultralow temperatures. Moreover, we demonstrate low-field Feshbach tuning of the contact interaction in our strongly dipolar BEC.

Our experimental procedure to create a BEC of Er follows a simple and straightforward scheme, inspired by work on Yb atoms \cite{Takasu2003ssb,Fukuhara2007bec}. Our starting point is the narrow-line {\em yellow} magneto-optical trap (MOT) described in our very recent work \cite{Frisch2012nlm}; it operates on the 583-nm line (natural linewidth $190$~kHz). We choose this approach because narrow-line MOTs permit  to obtain samples  with a large number of atoms at temperatures in the lower microkelvin region. This allows a  direct and efficient transfer of atoms into optical dipole traps without the need for additional cooling stages \cite{Takasu2003ssb,Fukuhara2007bec}.  Our MOT gives
about $10^{8}$ atoms  at a temperature of $\unit[15]{\mu {\rm K}}$ \cite{note2}.

An additional very advantageous feature of our approach is that the MOT light automatically pumps the atoms into the lowest Zeeman sublevel $m_J=-6$, where $m_J$ is the projection quantum number of the total electronic angular momentum $J=6$. This effect results from the interplay between gravity and weak radiation pressure, which leads to a spatial down shift with respect to the zero of the magnetic quadrupole field \cite{Katori1999mot} and thus to a preferential absorption of  the vertical MOT beam with $\sigma^-$ polarization \cite{Frisch2012nlm}. The polarization of the sample is confirmed by Stern-Gerlach-type measurements.
\begin{figure}[t]
\includegraphics[width=0.9\columnwidth] {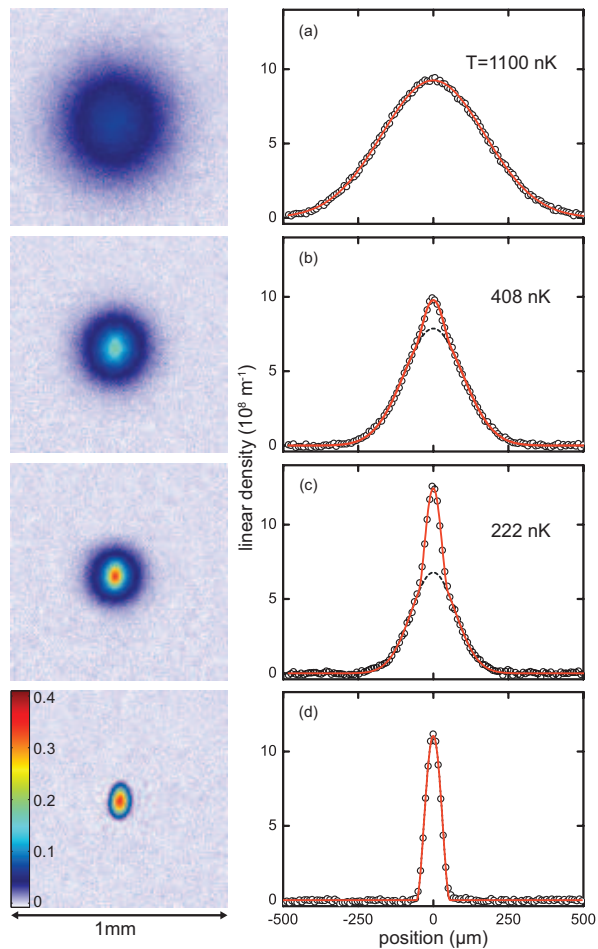}
\caption{(color online). Absorption images and integrated density profiles showing the BEC phase transition for different evaporation times.  The absorption images are an average of four images taken after $\unit[24]{ms}$ of expansion. The color bar shows the optical density. The solid lines are fits to the data using Gaussian (a), bimodal (b) and (c), and Thomas-Fermi (d) distribution. The dotted lines represent the  Gaussian part of the bimodal fit, describing the thermal atoms. From the fit we extract: $N=3.9\times 10^5$, $T=\unit[1100]{nK}$ (a), $N=2.1\times 10^5$, $T=\unit[408]{nK}$ (b), $N=1.6\times 10^5$, $T=\unit[222]{nK}$ (c), $N=6.8\times 10^4$ (d), where $N$ is the total atom number. For (b) and (c), we extract a condensate fraction of $5\%$ and $20\%$, respectively.}
\label{fig:Fig1}
\end{figure}

Our optical dipole trap geometry follows the concepts originally applied in experiments with Yb BEC \cite{Takasu2003ssb}.
The trap is created by crossing a tightly confining horizontal beam ($y$-axis) and a less focused vertical  beam ($z$-axis). The basic idea is that initially the atoms are predominantly trapped by the horizontal beam, whereas the vertical beam provides confinement relevant in the final stage of evaporation. The horizontal beam is derived from a 100-W broadband Yb fiber laser operating at $\unit[1075]{nm}$ and has an initial power of $\unit[10]{W}$. The beam has  an elliptic cross section with a waist of $\unit[30(40)]{\mu m}$ along the vertical (horizontal) direction. The vertical beam is produced by a 10-W Yb fiber laser source at $\unit[1064]{nm}$ and has an initial power of $\unit[8]{W}$. The beam profile is elliptic with a waist of $\unit[55(110)]{\mu m}$ along (perpendicular to) the axis of the horizontal beam.

We load the dipole trap during the MOT compression phase. We observe that the time period in which the compressed MOT and dipole trap coexist  is crucial for efficient loading. The number of atoms in the optical dipole trap exponentially approaches its maximum value with a time constant of about  $\unit[150]{ms}$.  After $\unit[600]{ms}$ of loading, we turn off the MOT beams and the quadrupole magnetic field, and we switch on a weak homogeneous magnetic field of about $\unit[300]{mG}$ along the vertical direction to preserve the spin polarization of the sample. We obtain $2.6 \times 10^6$ atoms at a temperature of  $42 \mu {\rm K}$ in the optical dipole trap. At this point, the atoms are mainly trapped by the horizontal beam. We measure oscillation frequencies $(\nu_x,\nu_y,\nu_z)=(1.3,0.016,1.95)$~kHz; the potential depth is estimated to be $\unit[560]{\mu K}$. The peak density and the peak phase-space density are  $\unit[1.7 \times 10^{13}]{cm^{-3}}$ and $1.6\times 10^{-4}$, respectively. These are our starting conditions for the evaporative cooling process.

Forced evaporative cooling is performed by reducing the power of the trapping beams in a nearly exponential ramp. The overall evaporation sequence has a duration of $\unit[5.5]{s}$ \cite{note3}.
We then turn off the trapping beams and let the atomic cloud expand  before applying standard absorption imaging. For imaging, we illuminate the atomic cloud with a $50$-${\mu {\rm s}}$ probe beam pulse \cite{note4}. The probe beam propagates horizontally at an angle of  $14^{\circ}$  with respect to the propagation axis ($y$-axis) of the horizontal trapping beam.

The phase transition from a thermal cloud to BEC manifests itself in a textbook-like bimodal distribution in the time-of-flight absorption images. Figure \ref{fig:Fig1} shows the absorption images and the corresponding linear density profiles for different final temperatures, i.\,e.\,for different stages of the evaporation. At higher temperatures the atomic distribution is thermal with the expected Gaussian profile resulting from the Maxwell-Boltzmann distribution; see Fig.\,\ref{fig:Fig1}(a).  By  cooling the atomic sample below the critical temperature, we clearly observe that the atomic density distribution has a bimodal profile with  a narrower and denser peak at the center, which represents the BEC (b). By further evaporating the sample, the  BEC fraction continuously  increases (c) until the thermal component is not anymore discernible and an essentially pure BEC  is formed with $7\times 10^{4}$ atoms (d).

To analyze our data, we fit a  bimodal distribution to the integrated time-of-flight absorption images. This distribution consists of a Gaussian function, which accounts for the thermal atoms, and an inverted integrated parabolic function for the BEC component in the Thomas-Fermi limit. Just after the onset of quantum degeneracy (BEC fraction $\sim 5\%$), we measure trap frequencies of $(\nu_x,\nu_y,\nu_z)=(208,70,299)$~Hz, atom number of $N=2.1\times 10^{5}$, and a temperature of $T=\unit[408]{n{\rm K}}$.
The critical temperature of $\unit[417]{n{\rm K}}$ as calculated from standard BEC theory  (without interaction shift) is consistent with this observation.

The evaporation efficiency is found to be remarkably high as 3.5 orders of magnitude in phase-space density are gained by losing a factor of ten in atom number.  This observation already points to favorable scattering parameters of the $^{168}$Er isotope. First  evaporative cooling experiments on the most abundant $^{166}$Er isotope  reveal a lower  efficiency in the final stage of evaporation, suggesting that a different strategy  might be needed to reach BEC.

To gain insight into the ultracold collisional properties of erbium we perform Feshbach spectroscopy \cite{Chin2010fri}  at low magnetic fields.
This measurement is done in a way that  allows us to identify both  the poles and zero crossings of Feshbach resonances \cite{Jochim2002mfc,Zaccanti2006cot}.
The basic idea here is to prepare the system at a variable  target value of the magnetic field and then to rapidly ($\unit[50]{ms}$) decrease the depth of the optical dipole trap by almost a factor of two. The sample stays near  thermal equilibrium with an effective temperature of  $\unit[2.2]{\mu {\rm K}}$ but features  a truncated energy distribution. We then let the system evolve at a constant trap depth for  $\unit[250]{ms}$, during which plain evaporative cooling and inelastic losses can occur depending on the scattering length. We finally switch off the trap and take time-of-flight images to determine the temperature and number of the atoms. The measurement is then repeated for variable magnetic field values.
Such a Feshbach scan shows resonance poles as loss features and  zero crossings as temperature maxima.

\begin{figure}[t]
\includegraphics[width=0.9\columnwidth] {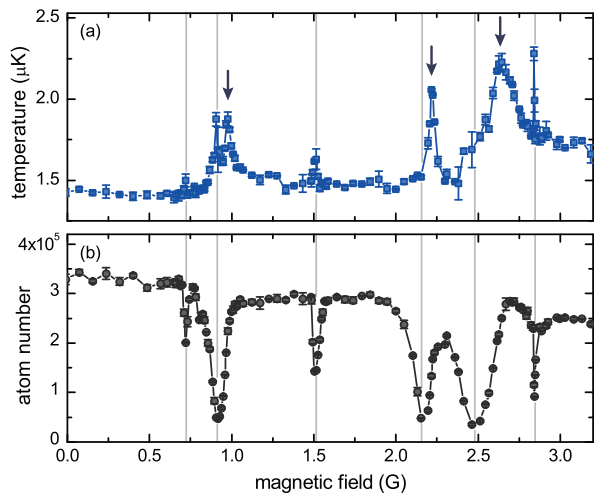}
\caption{(color online). Observation of  Feshbach resonances in Er-Er collisions. The measured temperature (a) and atom number (b) are plotted as a function of the magnetic field. The minima in the atom number indicate the Feshbach resonance poles, marked by the thin vertical lines. The maxima in the temperatures to the right of the three stronger loss features (arrows) are attributed to the respective zero crossings of the scattering length. The varying background in the atom number is presumably due to ramping effects caused by the sweep of the magnetic field across the resonances.}
\label{fig:fig2}
\end{figure}

Figure \ref{fig:fig2} shows the loss spectrum and the corresponding temperatures in the low magnetic-field range up to $\unit[3.2]{G}$ \cite{note5}.
Already in this narrow magnetic field range, the loss spectrum is very rich. We identify six pronounced resonant minima in the atom number that we interpret as being caused by Feshbach resonances.  For convenience, we determine the resonance positions with Gaussian fits, yielding $0.72$, $0.91$, $1.51$, $2.16$, $2.48$, and $\unit[2.85]{G}$.
The loss features show different strengths and widths.
For the three broader resonances at $0.91$, $2.16$, and $\unit[2.48]{G}$, we also observe the appearance of temperature maxima to the right of the loss minima (arrows in Fig.\, \ref{fig:fig2}). These temperature maxima mark the zero crossings of the scattering length. The other loss features are too narrow to provide clear signatures of the zero crossing.
From the difference in positions between the minima in the atom number and the maxima in temperature we estimate the widths $\Delta$ of the resonances. We find $\Delta=65$, $60$, and $\unit[180]{mG}$ for the resonances at $0.91$, $2.16$, and $\unit[2.48]{G}$, respectively.
\begin{figure}[t]
\includegraphics[width=0.9\columnwidth] {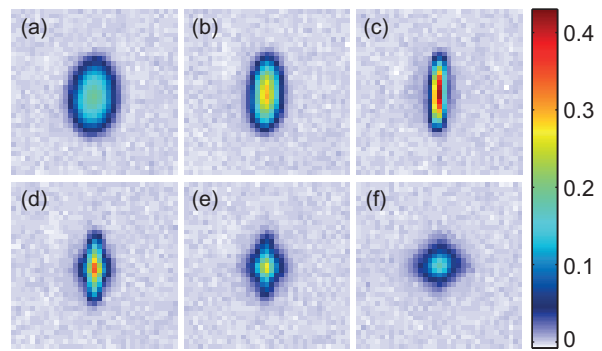}
\caption{(color online). Absorption images showing the $d$-wave collapse of the BEC. The field of view is $290 \mu {\rm m} \times 290 \mu {\rm m}$. The images are an average of eight pictures. The color bar shows the optical density. The images are taken for different target values of the magnetic field: $\unit[1.208]{G}$ (a), $\unit[0.963]{G}$ (b), $\unit[0.947]{G}$ (c), $\unit[0.942]{G}$ (d), $\unit[0.939]{G}$ (e), and $\unit[0.934]{G}$ (f). We note that the actual magnetic-field value might be slightly higher ($\unit[10]{mG}$)  because of  ramping issues and eddy currents \cite{Lahaye2008dwc}.}
\label{fig:fig3}
\end{figure}

In a strongly dipolar atomic gas,  universal dipolar scattering is present \cite{Roudnev2009uru, Bohn2009qud,Ni2010dco}, so that the total cross section for elastic scattering does not vanish at the zero crossings of the scattering length. For Er, a minimum cross section $\sigma_{\rm dip} = 8 \pi (30a_0)^2$ results from universal dipolar scattering, where $a_0$ is the Bohr radius. The fact that we observe temperature maxima near the zero crossings suggests a dominant role of $s$-wave scattering and not of dipolar scattering.
Preliminary cross-dimensional thermalization measurements indeed point to a scattering length between $150$ and $\unit[200]{a_0}$.

The existence of Feshbach resonances at low magnetic fields makes  the manipulation of the contact interaction in the Er BEC very convenient and straightforward. As a proof-of-principle experiment, we explore the controlled $d$-wave collapse of the BEC, following the procedure successfully applied by the Stuttgart group \cite{Lahaye2008dwc}. We first produce a pure BEC by evaporative cooling at $\unit[1.2]{G}$, which is above the position of the first broad Feshbach resonance ($\unit[0.91]{G}$). Here we obtain $3 \times 10^4$ atoms in the BEC, indicating that forced evaporation at this magnetic field is slightly less efficient. We then ramp down the magnetic field within $\unit[2]{ms}$ to a variable target value and let the sample evolve for $\unit[2]{ms}$ before switching off the trap. 
The magnetic field is kept constant at its target value during the first stage of the expansion ($\unit[15]{ms}$), where the main dynamics happens. We then set the magnetic field along the $y$-axis and we image the atomic cloud after additional $\unit[11]{ms}$ of expansion. Our results are summarized in Fig.\ref{fig:fig3}, where we show time-of-flight absorption images for different values of the target magnetic field.
We observe a dramatic change in the shape of the condensate when the magnetic field is reduced towards the zero crossing of the scattering length. At the magnetic field of evaporation, the aspect-ratio of the cloud is close to the one observed at zero magnetic field; see Fig.\,\ref{fig:fig3}(a). By changing the magnetic field to lower target values, the BEC shows a more and more anisotropic shape  (b) and (c). Below a critical magnetic field value, the BEC  develops a complicated {\em cloverleaf} pattern (d)-(f), which is the striking signature of the $d$-wave collapse in a  dipolar BEC  \cite{Lahaye2008dwc}.

In conclusion, we have demonstrated the first BEC of Er atoms and the tunability of its inter-particle interaction via Feshbach resonances. Our scattering data provide first sensitive input to understand the complex collisional behavior of  submerged-shell atoms.
The observation of the $d$-wave collapse is very encouraging in view of future experiments dedicated to the rich phenomena expected in strongly dipolar quantum gases.

We are grateful to  F.\,Schreck, S.\,Stellmer, and B.\,Pasquiou (Innsbruck Sr team) for their continuous support. We thank the Yb teams in Kyoto and Tokyo for their advices, M.\,Springer for technical support, and E.\,Zupani\v{c} for contributions in an earlier stage of the experiments. This work is supported by the Austrian Ministry of Science and Research (BMWF) and the Austrian Science Fund (FWF) through a START grant under Project No.\,Y479-N20 and by the European Research Council under Project No.\,259435.

\bibliographystyle{apsrev}


\begin{thebibliography}{32}
\bibitem[{not({\natexlab{a}})}]{note1}
\bibinfo{note}{The Er electronic configuration is [Xe]$4f^{12}6s^2$. The
  electronic ground state is a ${^3}H_6$ state, meaning $L=5$ ($H$-state),
  $S=1$ and $J=6$. $L$, $S$, $J$ denote the quantum numbers for the orbital
  angular momentum, for the spin, and for the total electronic angular
  momentum, respectively. $^{168}$Er has zero nuclear spin ($I=0$) and
  consequently do not exhibit a hyperfine structure.}
\bibitem[{not({\natexlab{b}})}]{note2}
\bibinfo{note}{The MOT stage includes a $15$-${\rm s}$ loading phase, in which
  the light is red detuned by $\unit[-9.5]{MHz}$, and a compression phase, in
  which we reduce the detuning to $\unit[-0.8]{MHz}$ and the MOT beam intensity
  by about a factor of 400. The compression phase is used to obtain colder and
  denser samples and it is optimized for loading efficiency into the dipole
  trap.}
\bibitem[{not({\natexlab{c}})}]{note3}
\bibinfo{note}{Initially only the power of the horizontal beam is reduced to
  its final value of $\unit[230]{mW}$, while the vertical beam power is reduced
  during the last $\unit[600]{ms}$ of evaporation to its final value of
  $\unit[1.5]{W}$.}
\bibitem[{not({\natexlab{d}})}]{note4}
\bibinfo{note}{We use the strong blue transition at $\unit[401]{nm}$ for
  imaging.}
\bibitem[{not({\natexlab{e}})}]{note5}
\bibinfo{note}{The magnetic field is calibrated via rf-spectroscopy between
  Zeeman sublevels. From the calibration we estimate a short-term stability of
  $\unit[1]{mG}$ and a long-term stability of about $\unit[10]{mG}$.}
\end{thebibliography}

\end{document}